\documentclass[aps,pra,twocolumn,longbibliography,superscriptaddress]{revtex4-2}

\usepackage{dcolumn}
\usepackage{bm}
\usepackage{subfigure}
\usepackage[normalem]{ulem}
\usepackage{enumerate}
\usepackage{cancel}
\usepackage{flushend}
\usepackage{bbold}
\usepackage{mathtools}
\usepackage{float}
\usepackage{stmaryrd}
\usepackage{colortbl}
\usepackage[table]{xcolor}
\usepackage{array,ragged2e}
\usepackage{amssymb}
\usepackage{wasysym,graphicx,multirow,textcomp}
\usepackage{url}
\usepackage{romannum}
\usepackage{physics}

\usepackage{changes}

\usepackage[colorlinks = true,linkcolor = red,citecolor = magenta]{hyperref}
\usepackage[sort&compress]{natbib}

\usepackage{tikz-cd}

\usepackage{braket}
\usepackage{url}

\makeatletter
\newcommand{\thickhline}{%
    \noalign {\ifnum 0=`}\fi \hrule height 1pt
    \futurelet \reserved@a \@xhline
}
\newcolumntype{"}{@{\hskip\tabcolsep\vrule width 1pt\hskip\tabcolsep}}
\makeatother

\newcommand{\figsizeone}{0.9}

\newcommand{\shead}[1]{\textcolor{blue}{\it{#1}}}

\newcommand{\eq}[1]{Eq.~(\ref{#1})}

\begin{document}

\draft

\title{Non-orientable Exceptional Points in Twisted Boundary Systems}

\author{Jung-Wan Ryu}
    \altaffiliation{These authors contributed equally.}
    \address{Center for Theoretical Physics of Complex Systems, Institute for Basic Science (IBS), Daejeon 34126, Republic of Korea}
    \address{Basic Science Program, Korea University of Science and Technology (UST), Daejeon 34113, Republic of Korea}

\author{Jae-Ho Han}
    \altaffiliation{These authors contributed equally.}
    \address{Department of Physics, Korea Advanced Institute of Science and Technology (KAIST), Daejeon 34141, Republic of Korea}

\author{Moon Jip Park}
    \address{Department of Physics, Hanyang University, Seoul 04763, Republic of Korea}
    \address{Research Institute for Natural Science and High Pressure, Hanyang University, Seoul, 04763, Republic of Korea}

\author{Hee Chul Park}
    \email{hc2725@gmail.com}
    \address{Department of Physics, Pukyong National University, Busan 48513, Republic of Korea}

\author{Chang-Hwan Yi}
    \email{yichanghwan@hanmail.net}
    \address{Center for Theoretical Physics of Complex Systems, Institute for Basic Science (IBS), Daejeon 34126, Republic of Korea}

\date{\today}

\begin{abstract}
Non-orientable manifolds, such as the M\"obius strip and the Klein bottle, defy conventional geometric intuition through their twisted boundary conditions. As a result, topological defects on non-orientable manifolds give rise to novel physical phenomena. We study the adiabatic transport of exceptional points (EPs) along non-orientable closed loops and uncover distinct topological responses arising from the lack of global orientation. Notably, we demonstrate that the cyclic permutation of eigenstates across an EP depends sensitively on the loop orientation, yielding inequivalent braid representations for clockwise and counterclockwise encirclement; this is a feature unique to non-orientable geometries. Orientation-dependent geometric quantities, such as the winding number, cannot be consistently defined due to the absence of a global orientation. However, when a boundary is introduced, such quantities become well defined within the local interior, even though the global manifold remains non-orientable. We further demonstrate the adiabatic evolution of EPs and the emergence of orientation-sensitive observables in a Klein Brillouin zone, described by an effective non-Hermitian Hamiltonian that preserves momentum-space glide symmetry. Finally, we numerically implement these ideas in a microdisk cavity with embedded scatterers using synthetic momenta.
\end{abstract}

\flushbottom
\maketitle

\shead{Introduction} Topological defects, such as diabolic and exceptional point, are characterized by orientation-dependent invariants. For instance, non-Hermitian topological invariants has been categorized into three groups:
(i) integer winding numbers arising from the intrinsic complexity of non-Hermitian bands \cite{Yao18a, Gon18, Lee19, Gha19, Kaw19b, Wan21c, Din22}, (ii) fractional winding numbers resulting from the non-separability of non-Hermitian bands \cite{Ryu12, Lee16, Zho18, Pap18, Woj20, Li21, Hu21, Wan21b, Pat22, Koe23, Zho23, Ryu24}, and (iii) complex Berry phases associated with the connectivity of non-Hermitian eigenstates \cite{Gar88, Dat90, Mos99, Lia13}. Exceptional points (EPs), where both eigenvalues and eigenstates coalesce \cite{Kat66, Hei90, Hei04, Dem04, Gao15, Dop16, She18, Yoo18, Zha19, Che20, Tan20, Li20, Liu20, Yan21, Yu21, Shu22, Sch22}, can endow non-Hermitian bands and their eigenstates with topologically nontrivial fractional winding numbers and complex Berry phases. These topological properties are governed by homotopy equivalence theory and permutation groups \cite{Ryu12, Lee12, Zho18, Ryu22, Ryu24}. Moreover, the topology of eigenvalue trajectories offers crucial insights into physical phenomena beyond conventional Hermitian frameworks. A key mathematical tool for understanding these systems is braid theory, which describes how eigenstates or energy bands interchange under adiabatic parameter evolution. When EPs are encircled in parameter space, the eigenstates can undergo nontrivial braiding, classified by representations of the braid group \cite{Hel20, Gha20, Zha23b, Gur24}. Extending this framework, knot theory is employed to characterize the global entanglement of eigenvalue trajectories in multi-parameter spaces \cite{Hu21, Pat22, Yu22b}.

Non-orientable manifolds are geometric spaces where a global orientation cannot be consistently defined. Unlike orientable manifolds, where a continuous choice of orientation is possible throughout the entire space, non-orientable manifolds exhibit intrinsic topological constraints that prevent such a definition \cite{Hat01, Nak03, Lee10}. Canonical examples of non-orientable 2D manifolds include the M\"obius strip, where a closed path reverses orientation; the Klein bottle, a closed surface without boundary; and the real projective plane, constructed by identifying antipodal points on a sphere. These examples illustrate how non-orientable surfaces can emerge by imposing various boundary conditions, offering key insights into their roles in physical systems. In physics, the topology of the Brillouin zone is equivalent to the torus. However, by imposing certain symmetries such as glide-reflection symmetry, the reduced Brillouin zone can become non-orientable \cite{Che22, Fon24, She24, Zhu24}. Imposing these structures can serve as a testbed to investigate unique topological phenomena that are absent in orientable spaces.

\begin{figure}
\includegraphics[width=\figsizeone\linewidth]{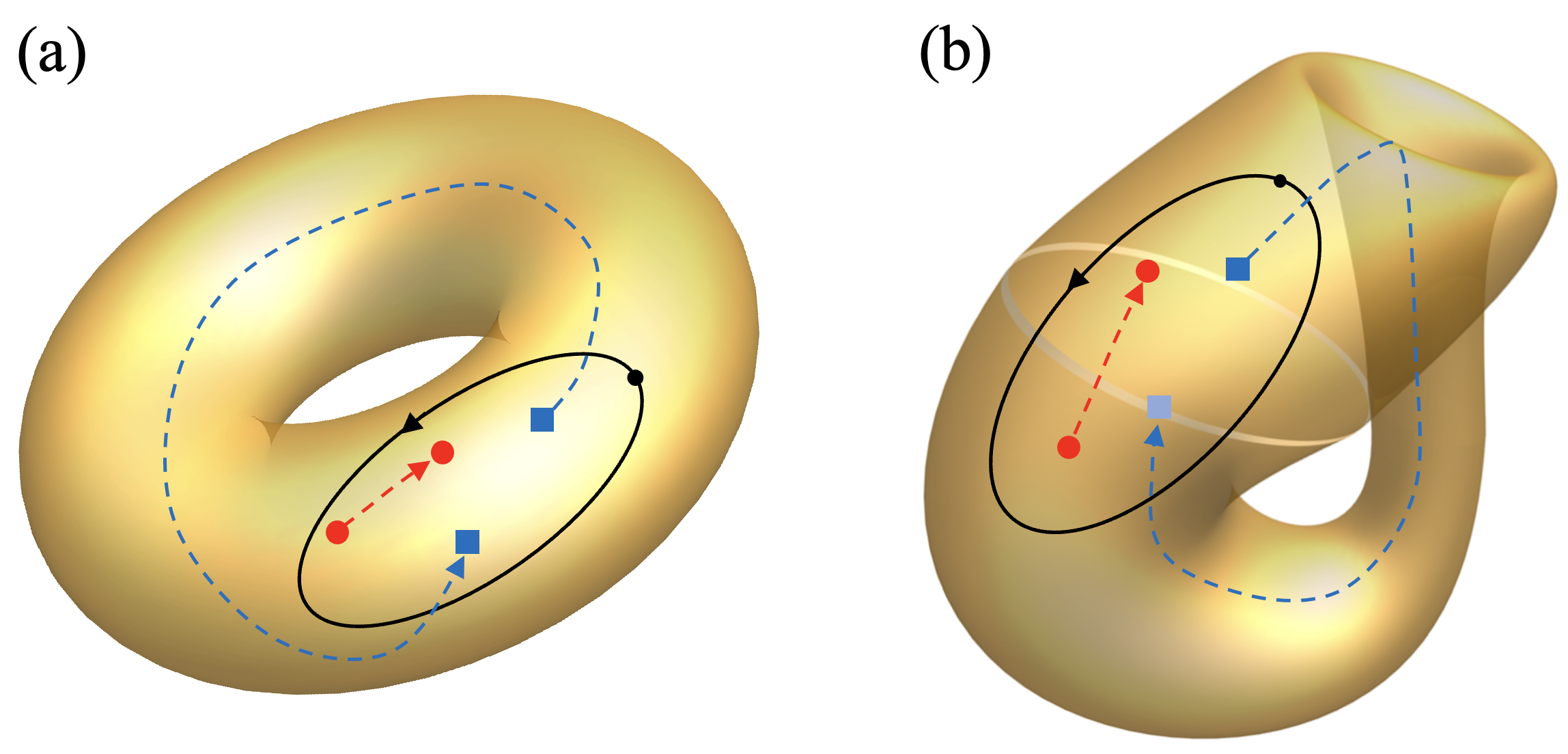}
\caption{Adiabatic evolutions (red and blue dashed arrows) of two EPs (red circles and blue rectangles) as system parameters are adiabatically varied on (a) a toroidal surface and (b) a Klein bottle surface. The light blue rectangle represents an EP located on the back side of the surface.}
\label{fig1}
\end{figure}

In this work, we explore the adiabatic evolution of EPs and establish an exceptional classification that characterizes the phases of eigenenergies and eigenstates in non-Hermitian bands, namely braids and complex Berry phases associated with homotopy equivalence classes, on non-orientable manifolds. Using the Klein bottle as an example (Fig.~\ref{fig1}), we illustrate how the classification of EPs differs from that on orientable manifolds. When the system parameters in the Hamiltonian are varied adiabatically, EPs can trace a closed loop in Brillouin zone and return to their original position. While the topological class remains unchanged on orientable manifolds, it can change on a non-orientable Brillouin zone even though the loop is closed. We also examine additional properties associated with EPs, such as Berry phases, vorticities, and state chirality, which may or may not change during such adiabatic loops in Brillouin zone. Finally, we propose a possible realization of non-orientable Brillouin zones and numerically demonstrate these phenomena in a microdisk cavity with multiple scatterers, leveraging synthetic momenta. These results deepen our understanding of non-Hermitian topology and offer new perspectives on topological classification of non-Hermitian Hamiltonians in non-orientable momentum spaces.

\shead{Exceptional Classification} When encircling EPs, the loop belongs to a specific homotopy equivalence class determined by the EPs enclosed within it. These classes reflect the braid group structure, which generally depends on the number and arrangement of EPs within the loop. However, in generic cases with only second-order EPs, one can focus on the two bands involved, where the topology and Berry phases are determined by the number of enclosed EPs, regardless of their specific configuration. A detailed classification of non-Hermitian bands in two band Hamiltonians based on EPs, using braid representations and Berry phases, including winding numbers, braid operations, and the relation to eigenstate evolution, is presented in the Appendix.

When EPs move out of the encircling loop, a phase transition occurs and the class of the loop changes. If the EP re-enters the loop, the class may return to its initial class. This behavior holds for orientable Brillouin zones, where the EP’s trajectory forms either a contractible or non-contractible loop [Fig.\ref{fig1}(a)]. However, this is not necessarily the case for non-orientable Brillouin zones [Fig.\ref{fig1}(b)]. In this work, we investigate the exceptional classification in which the topological invariants, namely the braid representations and the accumulated Berry phases, are defined through adiabatic evolutions along closed loops in a non-orientable Brillouin zone.

\begin{figure*}
\centering
\includegraphics[width=\figsizeone\linewidth]{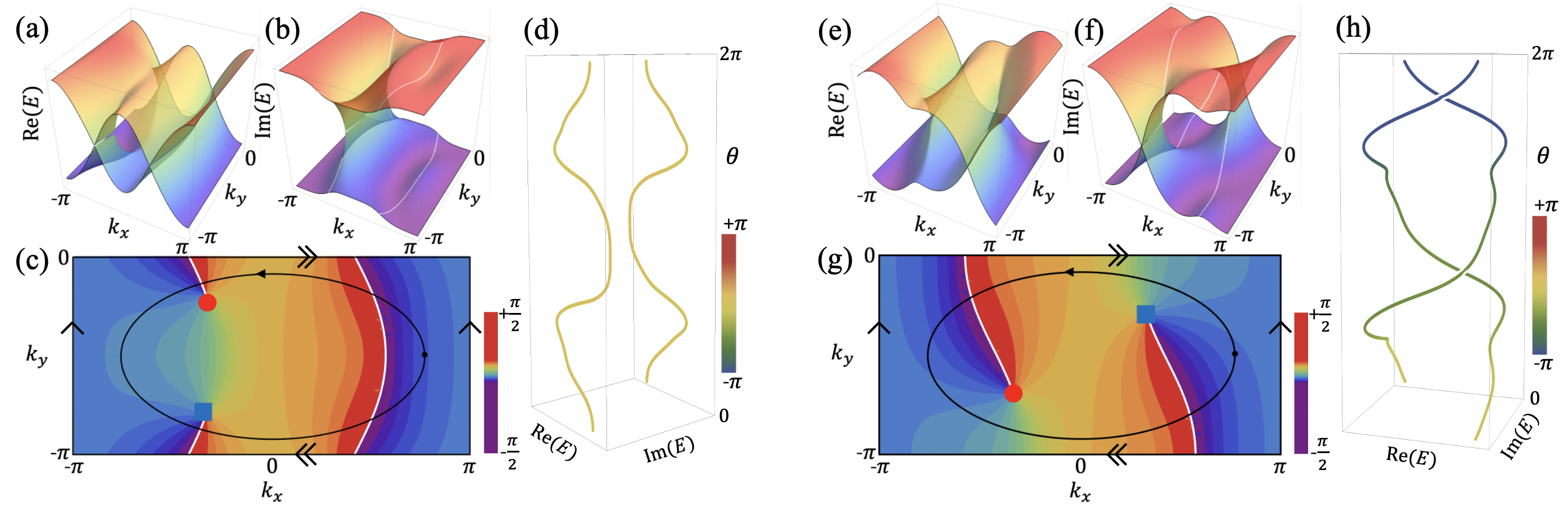}
\caption{Adiabatic evolutions of EPs in the KBZ and homotopy equivalence classes for $\gamma=0$ [(a)-(d)] and $\gamma=1$ [(e)-(h)] in the Hamiltonian, Eq.~(\ref{eq:Hamiltonian}). (a) Real and (b) imaginary parts of complex eigenenergies of the Hamiltonian with $\gamma=0$. (c) Two EPs (red circle and blue rectangle) and encircling loop (black closed loop) in the KBZ. The colors represents arguments of complex energy difference and the white lines are real branch cuts. The jump of $\pm \pi$ shows the direction of braid operation when the closed loop across the real branch cuts. (d) Trivial braids of two bands without accumulated Berry phases as $\theta$ increases on the closed loop. (e), (f), and (g) are the same as (a), (b), and (c), respectively, but with $\gamma=1$. (h) Non-trivial braids of two bands with accumulated Berry phases $\pi$ as $\theta$ increases on the closed loop.}
\label{fig3}
\end{figure*}

\shead{Non-Hermitian Hamiltonians and Klein Brillouin zone} 
Nonsymmorphic symmetries in crystalline materials have attracted considerable interest in the condensed matter community, owing to their ability to enforce nontrivial band topology and to facilitate unconventional superconductivity. In two-dimensional systems, glide reflection is the only relevant nonsymmorphic operation. Its implementation in momentum space yields a non-orientable manifold. Let us impose a momentum-space glide symmetry condition in two-dimensional \( (k_x, k_y) \) plane given by  
\begin{equation}
\label{eq:glide}
H(k_x, k_y) = H(-k_x, k_y + \pi),    
\end{equation}
which imposes a redundancy in momentum space \cite{Fon24}. This effectively modifies the topology of the Brillouin zone from a torus to a non-orientable fundamental domain described as a Klein bottle (\( K_2 \)) in the \( (k_x, k_y) \) plane. We call the Brillouin zone with Klein bottle topology as Klein Brillouin zone (KBZ). To describe this system, a two-band model is constructed using an effective Bloch non-Hermitian Hamiltonian of the form
\begin{equation}
\label{eq:Hamiltonian}
    H(\bold{k}) = d_x(\bold{k}) \sigma_x + d_y(\bold{k}) \sigma_y,
\end{equation}
where the vector \( \bold{d}(\bold{k}) = (d_x(\bold{k}), d_y(\bold{k})) \) defines the momentum-dependent coefficients subject to the glide symmetry constraint. The explicit parameterization is given by  
\begin{eqnarray}
d_x(\bold{k}) &=& \cos k_x + i \alpha, \\
d_y(\bold{k}) &=& - \sin k_x \left[(1-\gamma) \sin k_y + \gamma \cos k_y \right] - \frac{1}{2} + i \beta .    
\end{eqnarray}
The specific choice of these functions ensures that the system supports EPs at particular momentum-space locations where \( \|\bold{d}(\bold{k})\| = 0 \). 

\shead{Adiabatic evolutions of two EPs and braids.} We obtain two bands with complex energies when $\alpha = 1.0$ and $\beta = 0.5$, as a function of $\gamma$. The real and imaginary parts of the complex energies of the two bands show two EPs and real and imaginary branch cuts between them in the KBZ (Fig.~\ref{fig3}). We set an encircling loop including two EPs, with $k_x(\theta) = 2.5 \cos \theta$, $k_y(\theta) = 1.5 \sin \theta - \frac{\pi}{2}$, as the rotation angle $\theta$ increases from $0$ to $2\pi$. The homotopy equivalence class of the complex energy bands along the loop is determined solely by the combination of EPs inside the loop, according to exceptional classifications. If the loop contains the same combination of EPs, the braid class of the loop remains the same, regardless of the positions or order of the EPs. This is generally a local property inside the loop, independent of the global properties of the plane, such as the Brillouin zone. In Fig.~\ref{fig3}(d), the class is a trivial braid, $\tau^0$, which corresponds to an unlink in knot theory (see Fig.~\ref{fig2} in Appendix for details), and accumulated Berry phases of each band are zero, when $\gamma = 0$.

Under adiabatic variation of $\gamma$ from $0$ to $1$, two EPs persist, with only their positions changing in the KBZ. Figure~\ref{fig3} shows the EP configurations at $\gamma = 0$ and $1$. As shown, the EP marked in red moves downward while remaining inside the loop, whereas the blue one exits the loop. Consequently, the braid becomes $\tau^1$ or $\tau^{-1}$, corresponding to an unknot, as one EP remains inside the loop (see Fig.~S1 in Supplemental Material~\cite{SI}). The accumulated Berry phases of each loop are $\pi$ for double cyclic encricling of an EP. The blue EP eventually returns inside when $\gamma = 1$ and thus the loop encloses a non-trivial braid, $\tau^2$, corresponding to a Hopf link in knot theory and accumulated Berry phases of each bands are $\pi$ [Fig.~\ref{fig3}(h)]. If the two EPs at $\gamma = 1$ are the same as those at $\gamma = 0$, the braids at $\gamma = 0$ and $\gamma = 1$ must be identical, since the positions of the EPs do not influence the braid structure and accumulated Berry phases in the exceptional classification. The change in the braid reflects an alteration in the topological nature of the blue EP during the adiabatic evolution—an effect prohibited in a conventional toroidal Brillouin zone.

To quantify this topological change, we examine the vorticity defined near the EP. When the blue EP crosses the twisted KBZ boundary, its vorticity, which is the origin of the braid, changes sign. For a closed loop $\Gamma$ encircling the EP, the vorticity $\nu(\Gamma)$ is defined as follows \cite{Ley17, She18, Ryu24c}:
\begin{equation}
\nu (\Gamma) = - \frac{1}{2 \pi} \oint_{\Gamma} \nabla_{\mathbf{k}} \mathrm{arg}[E_{+} (\mathbf{k}) - E_{-} (\mathbf{k})] \cdot d \mathbf{k} . 
\label{eq:vorticity}
\end{equation}
Here, $\nu$ corresponds to the winding number of the eigenenergies $E_+$ and $E_-$ around the EP in the complex-energy plane. The vorticity $\nu$ of an EP is $\pm1/2$ for a closed loop $\Gamma$ encircling the EP. The $\pm$ signs indicate counterclockwise and clockwise rotations on the complex-energy plane, respectively. In non-Hermitian systems, vorticity is a topological invariant associated with the phase of eigenenergy, similar to the Berry phase from eigenstates. Since vorticity remains unchanged unless EPs are added or removed inside the loop, twice the total sum of the vorticities of EPs inside the loop equals the braid degree of the loop in two-band models. In Fig.~\ref{fig3}(g), two EPs have the same vorticities, which are impossible in a conventional toroidal Brillouin zone. As the parameter varies, two EPs can meet but they did not loose their defectiveness, unlike the merging of two EPs with the same vorticities make a vortex point which has no defectiveness and non-zero vorticity \cite{She18, Ryu24c}. In our model, an exceptional line emerges at $k_x = 0$ when two EPs merge (see Fig.~S2 in Supplemental Material~\cite{SI}). 

\shead{Non-orientability and orientation-dependent quantities} Now, let us consider the above situation on a Klein bottle surface [Fig.~\ref{fig1}(b)], which is a closed 2D surface without a boundary. When $\gamma = 0$, the two EPs are located on the front side of the Klein bottle surface. Mathematically, the two EPs and the encircling loop lie within the same orientable patch. As $\gamma$ increases, the blue EP escapes from the loop, while the red EP remains inside. Once the blue EP escapes, it becomes impossible to determine whether it is on the front or back side due to the non-orientability of the Klein bottle surface. However, when the blue EP returns inside the loop, it is then positioned on the back side. Namely, the blue EP travels along an orientaion-reversing path. If EPs remain inside the loop, the enclosed surface patch is effectively orientable. However, once EPs travel outside, the interior becomes non-orientable despite the vorticity signs being defined exactly. In contrast, on the Klein bottle surface, the vorticity sign outside the loop cannot be determined because of non-orientability.

It is worth noting that the vorticity is orientation-dependent, as its sign depends on the encircling direction, which can be verified from Eq.~(\ref{eq:vorticity}). In non-Hermitian systems, the braid and Berry phase are likewise orientation-dependent and can flip signs under adiabatic evolution on a non-orientable surface. Those quantities change their signs when EPs cross the twisted KBZ boundary, where such sign changes are well-defined due to the presence of the Brillouin zone boundaries. In contrast, no such identification of signs exists on the Klein bottle surface, where orientation cannot be defined without boundary.

\begin{figure*}
\centering
\includegraphics[width=0.825\linewidth]{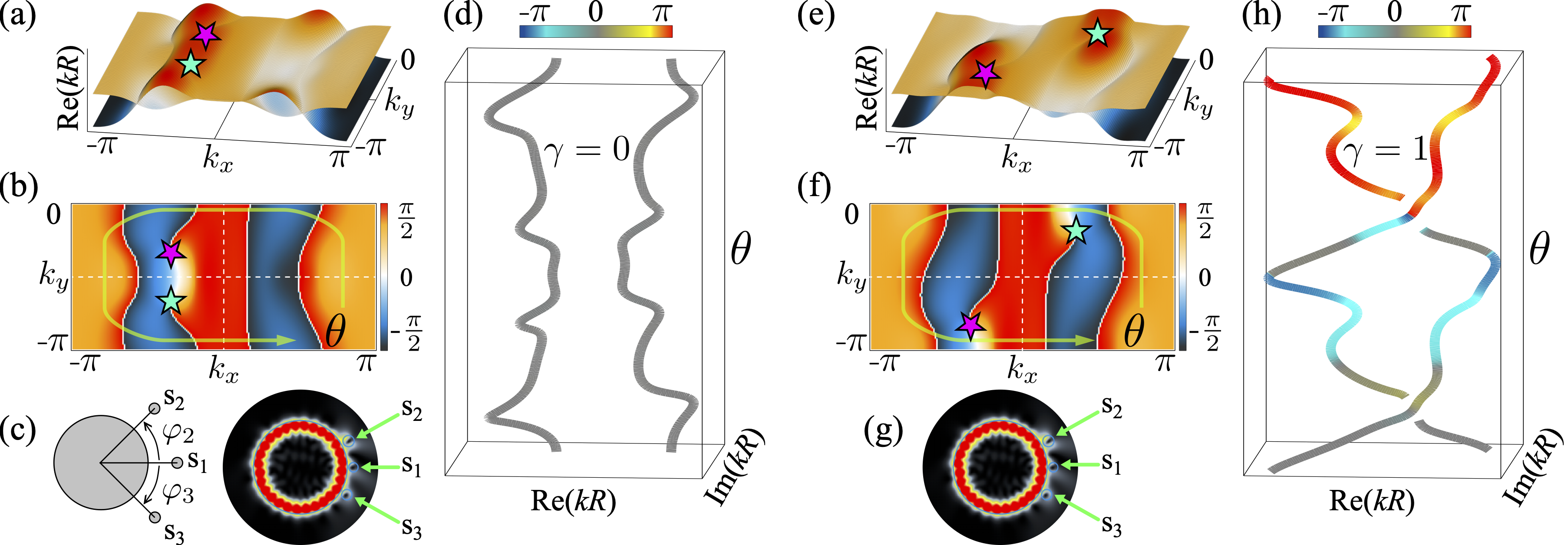}
\caption{Topological properties of energy bands of a dielectric disk with three scatterers for $\gamma=0$ [(a)-(d)] and $\gamma=1$ [(e)-(h)] in~\eq{eq:t3}. (a) Real parts of two selected energy bands obtained with $\gamma=0$ in the synthetic $(k_x, k_y)$ space, embedding two EPs (star symbols). (b) Arguments of the complex energy differences between the two bands. Red and blue colors indicate arguments of $\pm \pi/2$, while the white curves represent real branch cuts. The arrowed curve denotes the encircling path (with increasing $\theta$) used to examine state braiding and Berry phase accumulation along the closed loop. (c) The disk-scatterer system configuration with parameters $\varphi_i$ given in \eq{eq:t3} and spatial mode profiles $|\psi(\mathbf{r})|^2$ at the EPs of our dielectric disk system. The first scatterer $s_1$ is fixed, while the angular positions $\varphi_2$ and $\varphi_3$ of the other two scatterers $s_2$ and $s_3$ are varied with respect to the disk center. (d) Trivial braiding of the two bands, showing no Berry phase (colorbar) accumulation as $\theta$ evolves along the loop shown in (b). (e), (f), and (g) are the same as (a), (b), and (c), respectively, but with $\gamma=1$. (h) Non-trivial braiding of the two bands, showing $\pi$-Berry phase (colorbar) accumulation as $\theta$ evolves along the loop shown in (f).}
\label{fig4}
\end{figure*}

\shead{Implementation in a microcavity with synthetic momenta} While direct implementation of the KBZ in real physical systems remains challenging, recent progress has been made through tailored platforms based on synthetic dimensions \cite{Yua18}. One of the promising models in this direction is a one-dimensional photonic crystal with a synthetic momentum and lossy material \cite{Zho21, Zho23, Fon24, Ryu24b}. In this work, we propose an alternative model based on a microdisk with three scatterers whose positions are modulated as a function of synthetic momenta. This system provides an effective and conceptually transparent platform for theoretically exploring complicated topological properties of non-Hermitian eigenspace. Owing to their potential for ultra-sensitive sensing applications, microdisks with multiple scatterers have been widely studied and successfully demonstrated in high-Q whispering-gallery-mode experiments \cite{Wie14, Wie16, Pen16, Che17, Mao24}.

Given the dielectric microdisk with three scatterers [see Fig.~\ref{fig4}(c)], we obtain the optical eigenmodes by solving the two-dimensional Maxwell equations, which reduce to the Helmholtz equation, $-\nabla^2\vec{\psi}(\mathbf{r}) = n^2(\mathbf{r})k^2\vec{\psi}(\mathbf{r})$. For numerical computation, we employ the generalized Lorenz–Mie theory~\cite{Gag15,Gou11}, focusing on transverse-magnetic (TM) polarization [$\vec{\psi} = (0, 0, E_z)$], where the field $\psi$ and its normal derivative $\partial_\nu \psi$ are continuous across the cavity–vacuum interface~\cite{Jac99, Cha96}. As the pure-outgoing wave condition is imposed at infinity ($|\mathbf{r}| \to \infty$),  eigenfrequencies are complex-valued  $kR = \omega R / c\in \mathbb{C}$, where $|\text{Im}(kR)|$ is proportional to the decay rate. The refractive index $n(\mathbf{r})$ is piecewise constant: $n > 1$ inside the microdisk and $n_0 = 1$ outside.

Our synthetic momenta, $k_x$ and $k_y$, are defined such that the angular positions of the two scatterers, $s_2$ and $s_3$, with respect to the center of the disk, are given by
\begin{equation}
\begin{aligned}
\varphi_2 &= \frac{\pi}{7} + \frac{\pi}{28} \cos k_x, \\
\varphi_3 &= -\frac{\pi}{7} - \frac{\pi}{28} \sin k_x \left[(1-\gamma) \sin k_y + \gamma \cos k_y\right]
\end{aligned}
\label{eq:t3}
\end{equation}
The system preserves a momentum-space glide symmetry condition of Eq.~(\ref{eq:glide}) in two-dimensional synthetic momenta ($k_x$, $k_y$) plane. We examined $24$ complex energy bands, among which adjacent bands show various non-Hermitian two-band structures similar to the complex energy bands of effective Hamiltonians, Eq.~(\ref{eq:Hamiltonian}), with different parameters ($\alpha$, $\beta$) when $\gamma$ is between $0$ and $1$.

Two bands near $\mathrm{Re}(kR) \sim 5.36$ show a non-separable structure with two EPs (see Fig.~\ref{fig4} and Fig.~S3 in Supplemental Material~\cite{SI}). As $\gamma$ increases from $0$ to $1$, the adiabatic evolution of two EPs mimics those of the effective Hamiltonian model. During the adiabatic change of rotation angle $\theta$ from $0$ to $2 \pi$ on the encircling loop, the trival braids $\tau^0$ without Berry phases and non-trivial braids $\tau^2$ with Berry phases $\pi$ when $\gamma = 0$ and $\gamma = 1$, respectively, are shown in Figs.~\ref{fig4}(d) and (h). We note that there are many different topological structures of two adjacent bands in the systems resembling those of the two-bands Hamiltonian, Eq.~(\ref{eq:Hamiltonian}) (see Fig.~S3 in Supplemental Material~\cite{SI}).

The orientation-dependent quantities related to the EPs, e.g., braid on the closed loop, vorticity at an EP, and Berry phase of an eigenstate, change their signs when they across the KBZ boundary or the encircling loop. At EPs in asymmetric microcavities, the eigenstates are chiral optical modes, which have asymmetric angular momentum components (see Fig.~S4 in Supplemental Material~\cite{SI}) as well as the ultra-sensitivity for the external perturbation. The chirality of the eigenstate in the microcavity is then computed using its definition \cite{Wie11}: Chirality $\mathcal{R} = (|\alpha^+|^2 - |\alpha^-|^2)/(|\alpha^+|^2 + |\alpha^-|^2)$, where $|\alpha^{\pm}|$ are total sums of positive and negative angular momentum components, respectively. We obtain the spatial profiles of chiral eigenstates, as shown in Figs.~\ref{fig4}(c) and (g). However, the chirality of the optical modes at an EP does not change their signs when they across the twisted KBZ boundary, as the chirality of the mode is not an orientation-dependent quantity.

\shead{Summary}
We have proposed a classification framework for EPs in non-Hermitian systems defined on non-orientable manifolds, based on their adiabatic evolution in a momentum-based Hamiltonian with a KBZ. While the model is formulated in synthetic momentum space, we also propose a microcavity based photonic platform that comprises a whispering gallery mode resonator with embedded scatterers and captures the essential ingredients for realizing such EP structures. In this setting, we track changes in braid structure, vorticity, and Berry phase as EPs cross orientation-reversing boundaries, and identify distinct topological responses tied to physically relevant observables.

A key insight is that while quantities like vorticity and Berry phase are orientation-dependent and reverse sign across the KBZ boundary, the chirality of the optical mode at the EP remains invariant. This highlights the need to distinguish between orientation dependent topological invariants and robust, orientation independent observables, a distinction crucial for interpreting experiments in non-Hermitian systems.

\textit{Note added.} While this paper was in preparation, a related study \cite{Koe25} appeared on arXiv, focusing on the global structure of exceptional topology from a braid-theoretic perspective, uncovering EP fusion and fermion doubling anomalies. Our work complements this by focusing on the exceptional classification arising from explicit EP evolution in a momentum based Hamiltonian, and by proposing a realistic photonic setting in which the essential features of non-orientable exceptional topology, such as chirality, braiding, and adiabatic motion of EPs through orientation reversing boundaries, can be directly realized. Together, these two approaches offer a more complete picture of exceptional topology on non-orientable manifolds. We expect the interplay between exceptional topology and non-orientable geometry to emerge as a fertile direction for future theoretical and experimental research.

\section*{Acknowledgements}

We acknowledge financial support from the Institute for Basic Science in the Republic of Korea through the project IBS-R024-D1.

\section*{Appendix}

\subsection*{Exceptional classification in two-band Hamiltonians}

\begin{figure}
\centering
\includegraphics[width=1.0\linewidth]{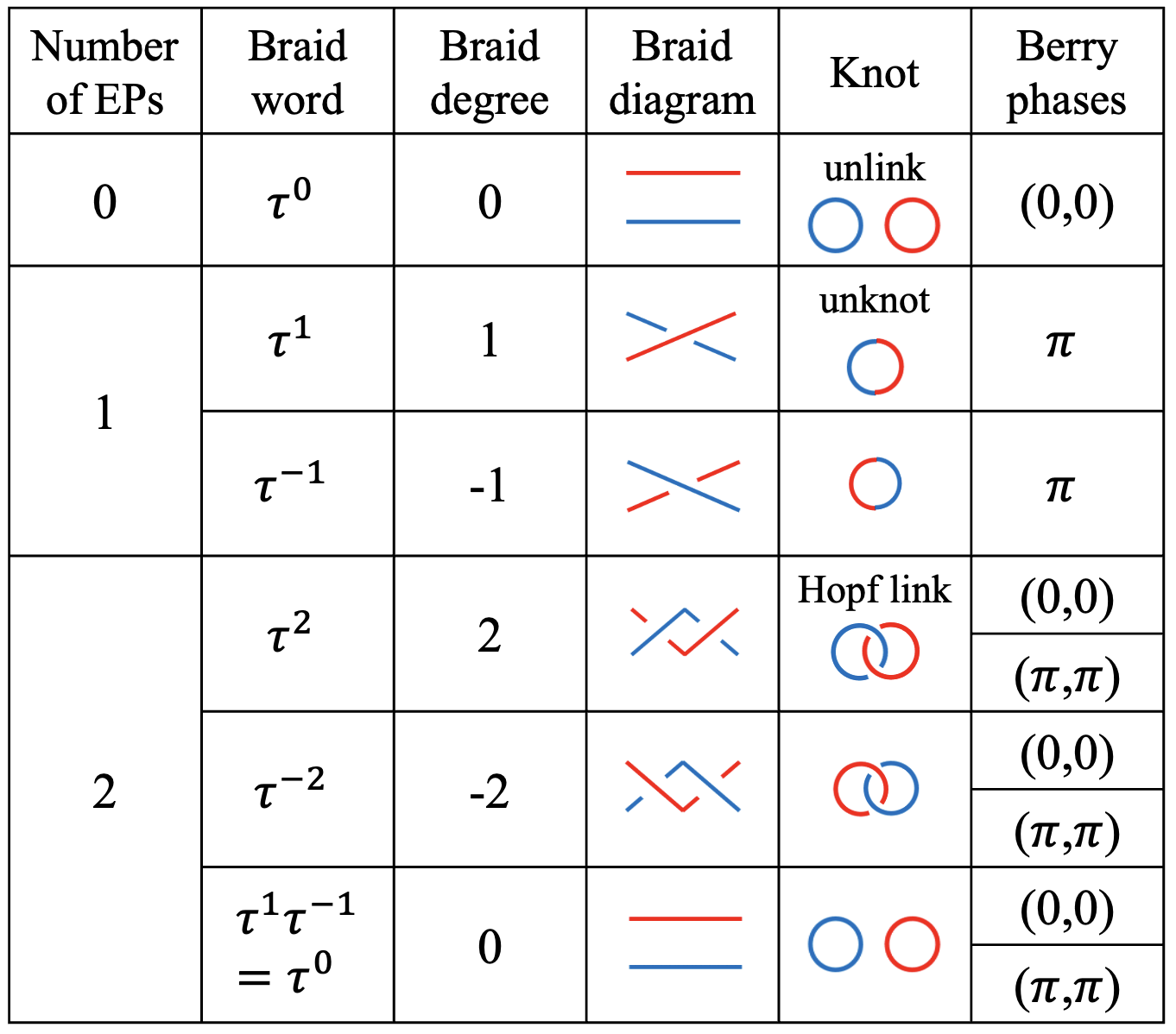}
\caption{Exceptional classification for braid group and Berry phases associated with the number of EPs inside a closed loop in two-band Hamiltonian with multiple EPs. Braid operators, braid degrees (winding numbers), braid diagrams, knots, and Berry phases depending on the number of EPs inside a loop.}
\label{fig2}
\end{figure}

In a two-band Hamiltonian, if an adiabatic trajectory forms a closed loop around an EP, the loop can be represented by braids, denoted as $\tau^1$ or its inverse $\tau^{-1}$, corresponding to an unknot in knot theory (Fig.~\ref{fig2}). After one complete cycle, the trajectory returns to its initial state through a specific braid transformation, which characterizes the permutation of eigenvalues along with their phase evolution, as well as the exchange of eigenstates. The topological characteristics of these braids are described by their winding numbers. The winding number (braid degree) $w$ characterizes how the eigenvalues evolve as the system encircles an EP and is mathematically defined as \cite{Gon18}:

\begin{equation}
    w = \sum_{n=1}^{N} \int_{-\pi}^{\pi} \frac{dk}{2\pi} \frac{d}{dk} \arg E_n(k),
\end{equation}
where \( E_n(k) \) are the complex eigenvalues of the $N$-band Hamiltonian \( H_N (k) \), and \( \arg E_n(k) \) represents their phase. This winding number determines the topological classification of the loop, indicating how the phases of the eigenvalues accumulate as momentum $k$ varies.

From this definition, the winding numbers of $\tau^1$ and $\tau^{-1}$ in two-band Hamiltonian ($N=2$) are $+1$ and $-1$, respectively. If no EP is enclosed within the loop, the braid $\tau^0$ is trivial, corresponding to an unlink, with a winding number of 0.

Now, consider the case where two EPs are inside the loop (black loops in Fig.~\ref{fig1}). This scenario gives rise to three distinct classes. After one full cycle, the two states return to their initial states with winding numbers $w = \pm 2$ for $\tau^{\pm 2}$, corresponding to Hopf links, or $w = 0$ for $\tau^1 \tau^{-1} = \tau^0 = I$, where $I$ is the identity element representing no braid. Even if the two EPs are repositioned within the loop, the classification remains unchanged. Another topological invariants related to the EPs are Berry phases accumulated during adiabatic parameter changes on the closed loop. The complex Berry phase \cite{Lia13, Gar88, Dat90, Mos99} can be defined as 
\begin{equation}
    \gamma = i \oint_{\mathcal C(\lambda)} \frac{\left< \phi (\lambda) | \partial_\lambda \psi (\lambda)\right>}{\left< \phi (\lambda) | \psi (\lambda)\right>} d\lambda.
\label{Berryphase}
\end{equation}
Here, $\mathcal C(\lambda)$ is a closed path in parameter space of the Hamiltonian, parameterized by $\lambda \in [0,1]$ with $\mathcal C(0) = \mathcal C(1)$, and $\left|\phi\right>$ and $\left|\psi\right>$ are left and right eigenstate of the Hamiltonian, respectively. The accumulated phases can be $(0,0)$ and $(\pi,\pi)$ for two separated bands depending on the braids (Fig.~\ref{fig2}), while $\pi$ for non-separated bands. The relations between braids and Berry phases in the case of a pair of EPs have four different cases depends on the Hamiltonians (see Supplemental Material~\cite{SI}). In the case of more than two-bands, different braids, $\tau_i$, and corresponding combinations of accumulated Berry phases of non-separable bands should be considered.

\clearpage

\begin{widetext}
\renewcommand{\theequation}{S\arabic{equation}}
\renewcommand{\thefigure}{S\arabic{figure}}
\setcounter{equation}{0}
\setcounter{figure}{0}

\section*{Supplemental Material for ``Non-orientable Exceptional Points in Twisted Boundary Systems"}
\section{An EP inside the loop}
\label{appendix:anEPloop}

Figure~\ref{figS1} shows a single EP enclosed by a closed loop in the KBZ, leading to the braid class $\tau_1$ when $\gamma = 0.5$ in the effective non-Hermitian Hamiltonian.

\begin{figure}
\centering
\includegraphics[width=0.5\linewidth]{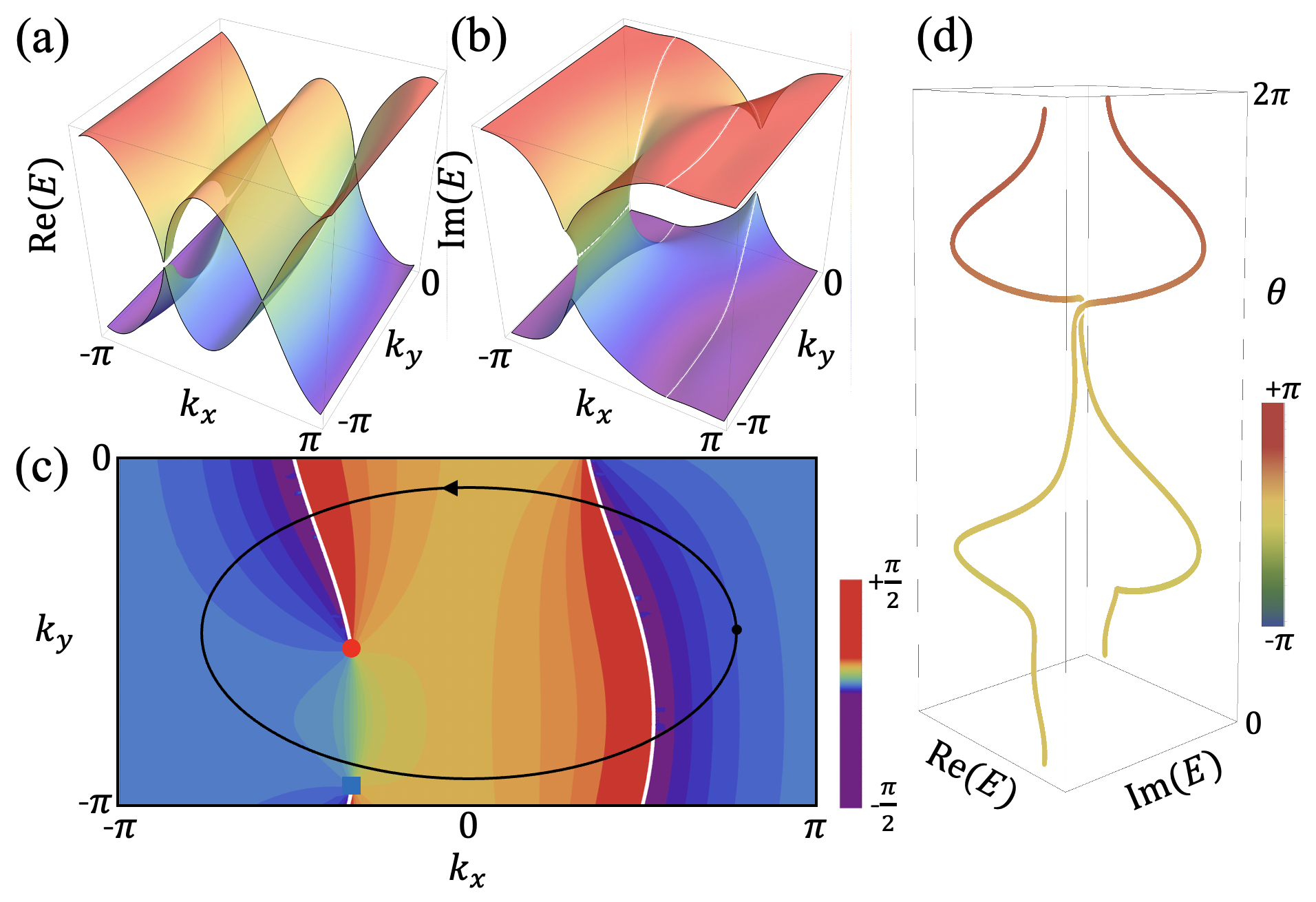}
\caption{Adiabatic evolutions of EPs in the KBZ and homotopy equivalence classes. (a) Real and (b) imaginary parts of complex eigenenergies of the Hamiltonian when $\gamma = 0.5$. (c) A encircling loop (black closed loop) including an EP (red circle) in the KBZ. The white lines represent real branch cuts. (d) Non-trivial braids of two bands with Berry phases as $\theta$ increases on the closed loop. The sum of accumulated Berry phases of two bands is $\pi$ as $\theta$ increases on the closed loop.}
\label{figS1}
\end{figure}

\section{Non-Hermitian bands in a Hamiltonian and a microcavity with synthetic momenta}
\label{appendix:bands}

\subsection{Non-Hermitian bands in a two-band Hamiltonian}

We examine topologically distinct band structures in the Hamiltonian at $\gamma = 1$ by varying the parameters $(\alpha, \beta)$. In the first case, two EPs with opposite vorticities approach and annihilate along the branch cuts, eliminating the imaginary branch cuts and leaving only real branch cuts in the KBZ (see Fig.~\ref{figS2}(a)). In the second case, two EPs with the same vorticities approach symmetric points, $(k_x, k_y) = (0, -\pi/2)$, where they form an exceptional line at $k_x = 0$ (see Fig.~\ref{figS2}(b)). The imaginary branch cut then transforms into a real branch cut through this line, resulting in only real branch cuts remaining (see Fig.~\ref{figS2}(c)).

\begin{figure}
\centering
\includegraphics[width=0.8\linewidth]{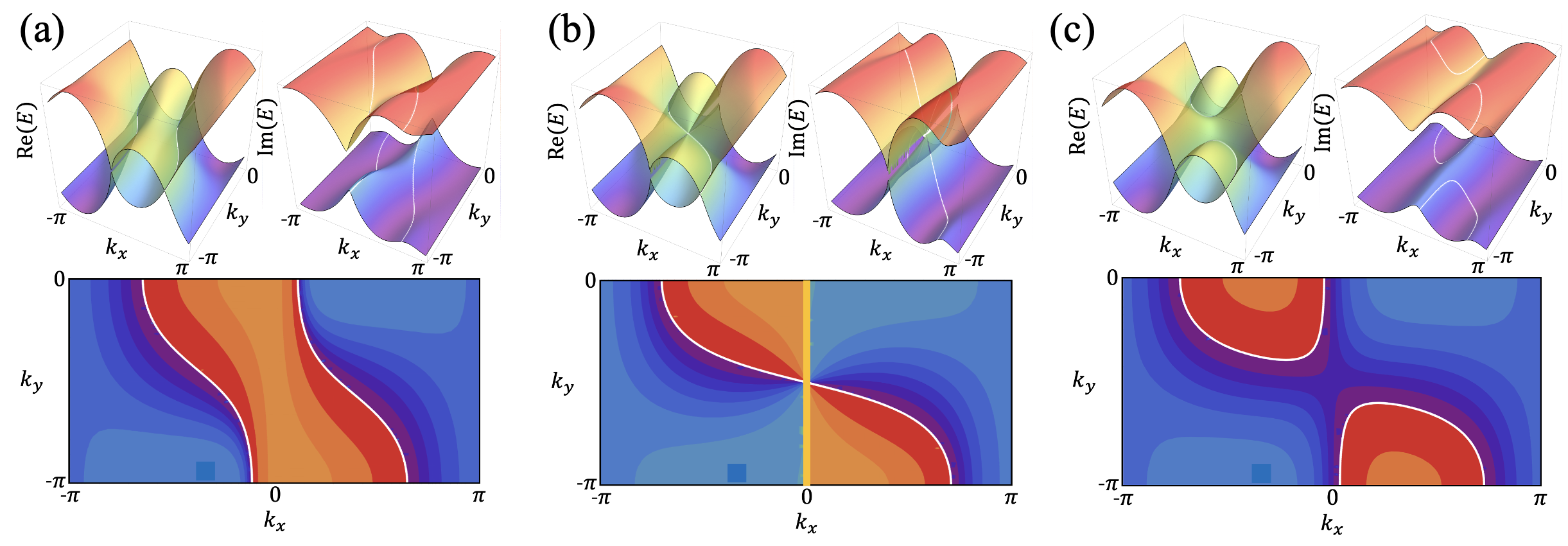}
\caption{Real, imaginary parts, and arguments of complex eigenenergies of the Hamiltonian when (a) $(\alpha,\beta,\gamma) = (0.9,1.0,1.0)$, (b) $(\alpha,\beta,\gamma) = (0.5,1.0,1.0)$, and (c) $(\alpha,\beta,\gamma) = (0.5,1.3,1.0)$. The orange line in (b) represent an exceptional line.}
\label{figS2}
\end{figure}

\subsection{Two adjacent bands in a microcavity with synthetic momenta}

We obtain the two pairs of bands which have topologically equivalent branch cut structures to the non-Hermitian bands in the effective Hamiltonian cases (see Fig.~\ref{figS3}). 

\begin{figure}
\centering
\includegraphics[width=1.0\linewidth]{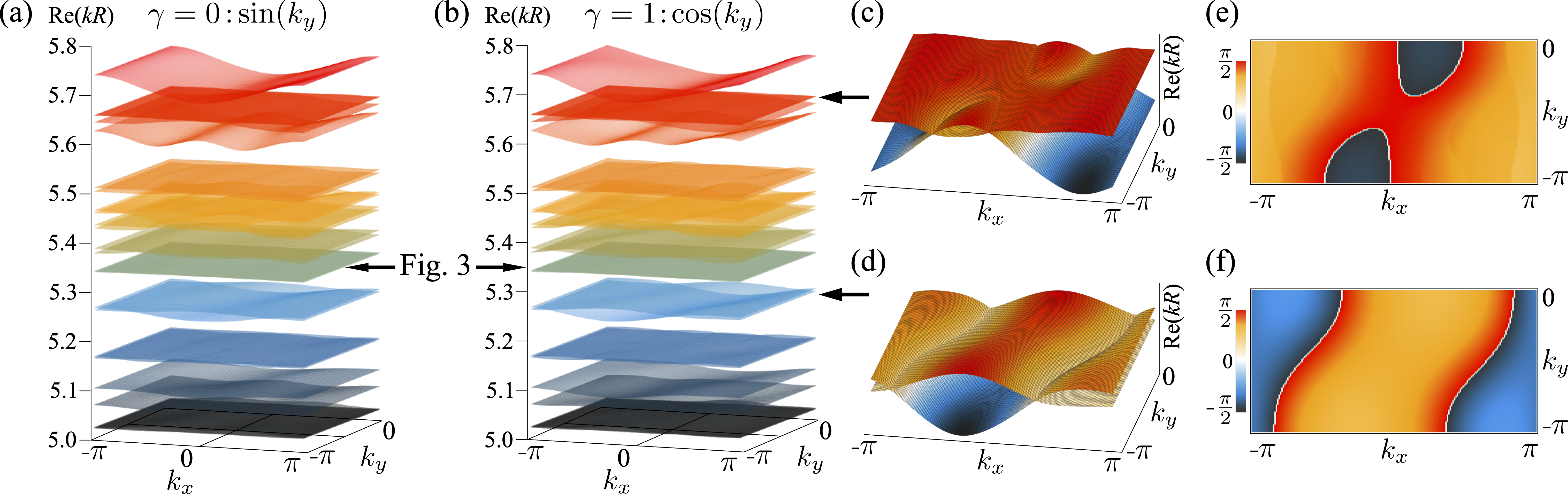}
\caption{The real part of 24 complex energy bands examined in the synthetic parameter space $(k_x,k_y)$ of a dielectric disk with three scatterers with (a) $\gamma=0$ and (b) $\gamma=1$ [cf. Eq. (6) in the main text]. The bands marked by arrows are the ones analyzed in Fig.~3 in the main text. (c) The selected bands (c) $nkR \sim 5.28$ and (d) $nkR \sim 5.69$ when $\gamma = 1$. The branch cut structures of the white curves in (e) and (f) are topologically equivalent to the ones in Figs.~\ref{figS2}(c) and (a), respectively.}
\label{figS3}
\end{figure}

\section{Chiral optical modes at EPs}
\label{appendix:chiral}

The chiral optical modes at EPs have asymmetric angular momentum components (see Fig.~\ref{figS4}).

\begin{figure}
\centering
\includegraphics[width=0.5\linewidth]{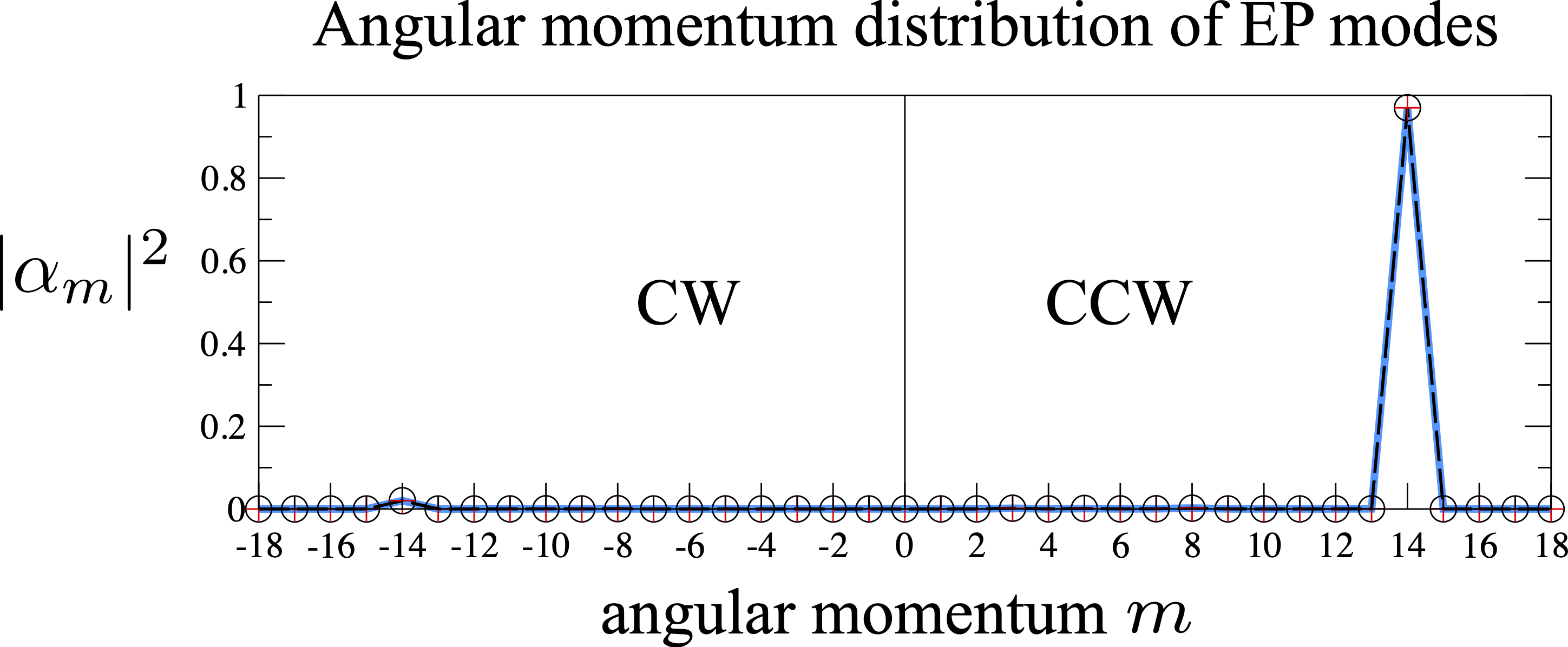}
\caption{The distribution of the expansion coefficient $\alpha_m$ of mode at EPs marked by stars in Fig.~3 in the main text; $\psi(\mathbf{r})=\sum_m\alpha_m J_m(nk\mathbf{r})e^{i m \theta}$. As we can confirm here, the EP modes exhibit almost perfect chirality, i.e., $\mathcal{R}\approx 1$.}
\label{figS4}
\end{figure}

\section{A pair of EPs: vorticity and Berry phase}
\label{appendix:EPpair}

We consider a closed loop that encircles two EPs. There are two types of EP pairs: counter-rotating EPs, which have opposite vorticities, and co-rotating EPs, which have the same vorticities. These are referred to as type-I and type-II EP pairs, respectively. The eigenenergy structures for type-I and type-II EP pairs are topologically equivalent to the energy structures given by $E_{\pm}^{\rm I} = \pm \sqrt{( z - z_1 )( z - z_2 )^{*}}$ and $E_{\pm}^{\rm II} = \pm \sqrt{( z - z_1 )( z - z_2 )}$, where $z_1$ and $z_2$ are the two EPs.
In the case of type-I EP pairs, we consider two different Hamiltonian which have the same energies. 
\begin{equation}
 \mathbf{Case~1:}~   H(z) = 
\begin{pmatrix} 
 0 & z-z_1 \\
 (z-z_2)* & 0 \\
\end{pmatrix} ,
\label{eq:model_1}
\end{equation}
where $z = x + i y$ is a two-dimensional parameter and there is an EP when $z=z_1$ or $z=z_2$. If $z_1$ and $z_2$ approach to the origin, the Hamiltonian can be rewritten as
\begin{equation}
    H(z) = 
\begin{pmatrix} 
 0 & z \\
 z^* & 0 \\
\end{pmatrix} .
\label{eq:model_1b}
\end{equation}
If we set $z=e^{i \theta}$, the eigenenergies and the eigenstates are 
\begin{equation}
    \lambda = -1, 1.
\label{eq:model_1b_ev}
\end{equation}
\begin{equation}
    \ket{\psi} = 
\begin{pmatrix} 
 -e^{i \theta} \\
 1 \\
\end{pmatrix} ,
\begin{pmatrix} 
 e^{i \theta} \\
 1 \\
\end{pmatrix}.
\label{eq:model_1b_es}
\end{equation}
The braid degrees are $0$ and accumulated Berry phases are $\pi$ during $\theta$ changes from zero to $2 \pi$. This is the same case that two EPs merge into a Dirac point. When encircling a pair of EPs in Eq.~(\ref{eq:model_1}), the results are the same.

\begin{equation}
 \bold{Case~2:}~   H(z) = 
\begin{pmatrix} 
 0 & 1 \\
 (z-z_1)(z-z_2)* & 0 \\
\end{pmatrix} .
\label{eq:model_2}
\end{equation}
If $z_1$ and $z_2$ approach to the origin, the Hamiltonian can be rewritten as
\begin{equation}
    H(z) = 
\begin{pmatrix} 
 0 & 1 \\
 |z|^2 & 0 \\
\end{pmatrix} .
\label{eq:model_2b}
\end{equation}
If we set $z=e^{i \theta}$, the eigenenergies and the eigenstates are 
\begin{equation}
    \lambda = -1, 1.
\label{eq:model_2b_ev}
\end{equation}
\begin{equation}
    \ket{\psi} = 
\begin{pmatrix} 
 -1 \\
 1 \\
\end{pmatrix} ,
\begin{pmatrix} 
 1 \\
 1 \\
\end{pmatrix}.
\label{eq:model_2b_es}
\end{equation}
The braid degrees are $0$ and accumulated Berry phases are $0$ during $\theta$ changes from zero to $2 \pi$. When encircling a pair of EPs in Eq.~(\ref{eq:model_2}), the results are the same.

In the case of type-II EP pairs, we also consider two different Hamiltonian which have the same energies. 
\begin{equation}
 \bold{Case~3:}~   H(z) = 
\begin{pmatrix} 
 0 & z-z_1 \\
 z-z_2 & 0 \\
\end{pmatrix} .
\label{eq:model_3}
\end{equation}
If $z_1$ and $z_2$ approach to the origin, the Hamiltonian can be rewritten as
\begin{equation}
    H(z) = 
\begin{pmatrix} 
 0 & z \\
 z & 0 \\
\end{pmatrix} .
\label{eq:model_3b}
\end{equation}
If we set $z=e^{i \theta}$, the eigenenergies and the eigenstates are 
\begin{equation}
    \lambda = -e^{-i \theta}, e^{i \theta}.
\label{eq:model_3b_ev}
\end{equation}
\begin{equation}
    \ket{\psi} = 
\begin{pmatrix} 
 -1 \\
 1 \\
\end{pmatrix} ,
\begin{pmatrix} 
 1 \\
 1 \\
\end{pmatrix}.
\label{eq:model_3b_es}
\end{equation}
The braid degrees are $2$ and accumulated Berry phases are $0$ during $\theta$ changes from zero to $2 \pi$. This is the same case that two EPs merge into a vortex point. When encircling a pair of EPs in Eq.~(\ref{eq:model_3}), the results are the same.

\begin{equation}
 \bold{Case~4:}~   H(z) = 
\begin{pmatrix} 
 0 & 1 \\
 (z-z_1)(z-z_2) & 0 \\
\end{pmatrix} .
\label{eq:model_4}
\end{equation}
If $z_1$ and $z_2$ approach to the origin, the Hamiltonian can be rewritten as
\begin{equation}
    H(z) = 
\begin{pmatrix} 
 0 & 1 \\
 z^2 & 0 \\
\end{pmatrix} .
\label{eq:model_4b}
\end{equation}
If we set $z=e^{i \theta}$, the eigenenergies and the eigenstates are 
\begin{equation}
    \lambda = -e^{i \theta}, e^{i \theta}.
\label{eq:model_4b_ev}
\end{equation}
\begin{equation}
    \ket{\psi} = 
\begin{pmatrix} 
 -e^{-i \theta} \\
 1 \\
\end{pmatrix} ,
\begin{pmatrix} 
 e^{-i \theta} \\
 1 \\
\end{pmatrix}.
\label{eq:model_4b_es}
\end{equation}
The braid degrees are $2$ and accumulated Berry phases are $\pi$ during $\theta$ changes from zero to $2 \pi$. When encircling a pair of EPs in Eq.~(\ref{eq:model_4}), the results are the same. The four cases are summarized in Table~\ref{tab01}.

\begin{table*}[ht]
\centering
    \begin{tabular}{l|cccc}
    \hline
        EP pairs & Case 1 & Case 2 & Case 3 & Case 4 \\
    \hline
        (i) Vorticities & ($\pm$,$\mp$) & ($\pm$,$\mp$) & ($\pm$,$\pm$) & ($\pm$,$\pm$) \\
        (ii) Braids & $\tau^0$ & $\tau^0$ & $\tau^{\pm 2}$ & $\tau^{\pm 2}$ \\
        (iii) Berry phases & $(\pi,\pi)$ & (0,0) & (0,0) & $(\pi,\pi)$ \\ 
        (iv) Merging & DP & Defective & VP & Defective \\
    \hline
    \end{tabular}
\caption{\label{tab01} Classification of EP pairs for two-band Hamiltonians. (i) Vorticities of each EP, (ii) braids and (iii) accumulated Berry phases during encircling the EP pairs, and (iv) phases when two EPs merge. DP and VP represent Dirac and vortex points, respectively, where there is no defectiveness.}
\end{table*}

\end{widetext}

\end{document}